# Probing fragile topology with a screw dislocation

Ying Wu[1, *], Zhi-Kang Lin[2, *], Yating Yang[3], Zhida Song[4, 5, 6, #], Feng Li[7, $], Jian-Hua Jiang[2, 8, †]

[1]*School of Physics, Nanjing University of Science and Technology, Nanjing 210094, China*

[2]*School of Physical Science and Technology, and Collaborative Innovation Center of Suzhou Nano Science and Technology, Soochow University, 1 Shizi Street, Suzhou 215006, China*

[2]*College of Mathematics and Physics, Beijing University of Chemical Technology, 100029, Beijing, China*

[4]*International Center for Quantum Materials, School of Physics, Peking University, Beijing 100871, China*

[5]*Hefei National Laboratory, Hefei 230088, China*

[6]*Collaborative Innovation Center of Quantum Matter, Beijing 100871, China*

[7]*Key Lab of Advanced Optoelectronic Quantum Architecture and Measurement (MOE), School of Physics, Beijing Institute of Technology, Beijing 100081, China*

[8]*Suzhou Institute for Advanced Research, University of Science and Technology of China, Suzhou 215123, China*

[*]These authors contributed equally to this work.

[#]Email: songzd@pku.edu.cn

[$]Email: phlifeng@bit.edu.cn

[†]Email: jianhuajiang@suda.edu.cn

**Fragile topology, akin to twisted bilayer graphene and the exotic phases therein, is a notable topological class with intriguing properties. However, due to its unique nature and the lack of bulk-edge correspondence, the experimental signature of fragile topology has been under debated since its birth. Here, we demonstrate experimentally that fragile topological phases with filling anomaly can be probed via screw dislocations, despite that they do not support gapless edge states. Using a designer hexagonal phononic crystal with a fragile topological band gap, we find that 1D gapless bound modes can emerge at a screw dislocation due to the bulk fragile topology. We then establish a connection between our system and the twisted boundary condition via the gauge invariance principle and illustrate that such an emergent**

**phenomenon is an intrinsic property of fragile topological phases with filling anomaly. We observe experimentally the 1D topological bound states using the pump-probe measurements of their dispersion and wavefunctions, which unveils a novel bulk-defect correspondence of fragile topology and a powerful tool for probing fragile topological phases and materials.**

*Introduction.* – In the past years, fragile topology [1-41] and its important role in the properties of twisted bilayer graphene [3, 23-27, 36], non-Abelian topology [28-34], flat bands [35-37] and related emergent phases have attracted much attention, and led to discoveries that connect band topology with correlated electron states [38-42] such as superconductivity [40-42]. Interestingly, the band topology can impose constraints on the superconducting order parameters as revealed by recent studies [40-42]. Fragile topological phases are characterized by nontrivial signatures in either the Wilson loops [5, 6, 13, 14] or the band representations [2, 5, 14, 16, 43, 44] but can be connected with some obstructed atomic insulator phases---insulators with the charge centers away from atomic centers---via adding some trivial bands [2]. It has been shown that fragile topological insulators do not support robust edge states, imposing challenges on their experimental signatures. Later, it was proposed that local gauge flux insertion [7] or twisted boundary conditions [45-50] can be used to probe the fragile topology. However, both the local gauge flux insertion and twisted boundary conditions are very challenging in genuine condensed matter experiments. Up till now, experimental signatures of fragile topology have not yet been observed in solid-state systems.

Here, we propose that screw dislocations can serve as an efficient probe of fragile topology. Since screw dislocations commonly exist in solid-state materials, it opens a new pathway toward the experimental identification of fragile topological phases. By using the arguments based on the dimensional reduction, we demonstrate that screw dislocations can be regarded as sources of local gauge flux [48, 49] and thus lead to the emergence of 1D gapless modes bound to the dislocations due to the intrinsic properties of the fragile topology. Equivalently, such an effect can also be connected to the twisted boundary condition via the gauge invariance principle. Therefore, the emergence of the 1D gapless bound states at a screw dislocation can be identified as a signature bulk-defect correspondence of fragile topological phases. We also remark that due to the rapid development of the field, the concept of fragile topology has been extended to more general cases, including the Euler insulators [29, 32-34, 51-53]. The investigation here is still constrained to the original fragile topological phases with nontrivial filling anomaly which is a major class of fragile topological phases that are of interest in many studies [1-27]. Our study reveals that for such fragile topological phases with filling anomaly, screw dislocations can serve as a powerful experimental tool for the probe of fragile topology.

*Fragile topological phononic crystals.* – We consider a fragile topological phase based on a bilayer construction. The system is illustrated in Fig. 1(a) in the tight-binding picture where two layers of honeycomb lattice (the nearest neighbor hopping is $t_1$) are coupled in a spiral way. We consider the situation with identical onsite energy at all sublattice sites. Therefore, the system is gapless without the interlayer couplings [Fig. 1(b)]. The spiral interlayer couplings $t_2$ open a band gap with nontrivial fragile topology [Fig. 1(c)], which can be revealed either by the nontrivial Wilson loop or the band representation analysis [Symmetry eigenvalues at the high symmetry points are given in both Figs. 1(b) and 1(c); see more details in Supplemental Materials].

The fragile topological band gap does not support robust gapless edge states. Instead, the edge states are often gapped as illustrated in Fig. 1(d). To yield truly topological responses, one needs to introduce a local gauge flux. We shall show here that such a local gauge flux can be introduced by the following procedures [48, 49]: First, we stack the 2D bilayer system periodically in the z direction to yield a 3D lattice system with vanishing coupling between the periodically repeated 2D bilayers. We then introduce a screw dislocation in the systems which breaks all the 2D lattice translation symmetry but keeps the periodic translation symmetry along the z direction. The introduction of a screw dislocation also provides a mechanism to connect all the previously separated 2D bilayers together. Finally, by performing the Fourier transformation in the z direction and regarding the system with a given $k_z$ as an effective 2D system, the effective 2D system carries a local artificial gauge flux $\phi = \vec{k} \cdot \vec{B}_v = k_z H$ at the screw dislocation where $\vec{B}_v = (0,0,H)$ is the Burgers vector of the screw dislocation and $H$ is the lattice constant along the $z$ direction [Fig. 1(e), see more details in Supplemental Materials]. With such a local artificial gauge flux, the fragile topology gives rise to the gapless bound states at the screw dislocation that traverse the entire topological band gap [Fig. 1(f)] according to the theories in Refs. [7, 45].

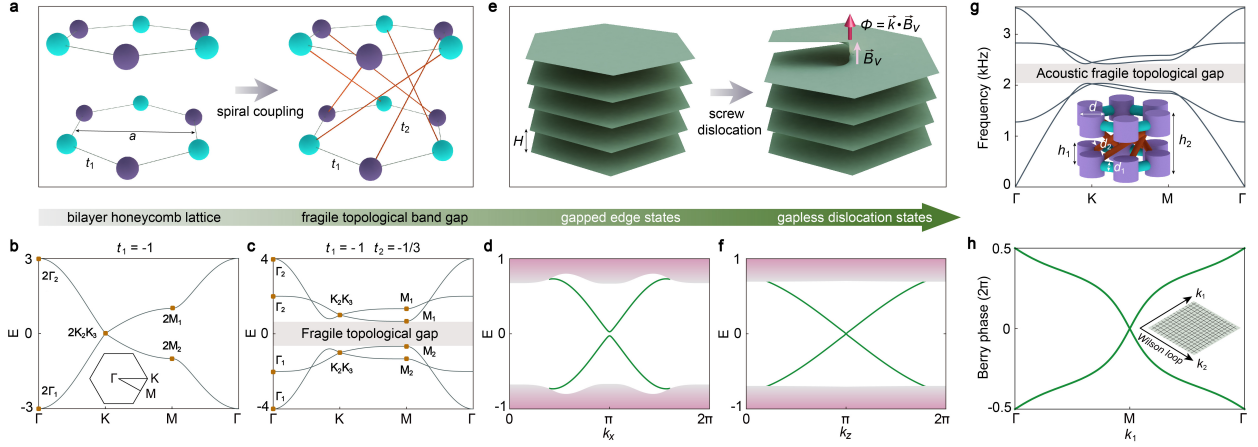

FIG. 1. (a) Illustration of the spirally coupled bilayer honeycomb lattice in the tight-binding representation. The black lines denote the nearest-neighbor hoppings $t_1$, while the red lines denote the spiral interlayer hoppings $t_2$. The lattice constant $a$ is set to unity throughout this study in tight-binding calculations. (b) Band structure of two decoupled honeycomb lattices with $t_1 = 1$. (c) Band structure of the spirally coupled bilayer honeycomb lattice with $t_1 = 1$ and $t_2 = 1/3$, which exhibits a fragile topological band gap. The symmetry representations of Bloch eigenstates at high-symmetry points $\Gamma$, $K$, and $M$ are labeled. The notation of symmetry representations can be referred to the character tables in Table. 1. (d) Band structure versus $k_x$ of a ribbon supercell with open boundary condition in the $y$ direction, which exhibits the gapped edge states that are the hallmark feature of the fragile topological band gap. (e) Schematic illustration of a screw dislocation that brings in a local gauge flux $\phi = \vec{k} \cdot \vec{B}_v$ with $\vec{B}_v$ denoting the Burger's vector. Here each plane represents a bilayer. (f) Gapless spectral flows versus $k_z$ induced by the screw dislocation. (g) Acoustic analog of the fragile band gap. The inset shows the acoustic unit cell, where the purple cylindrical cavities with a diameter of $d$ and a height of $h_1$ play the role of atomic sites. The intra- and inter-layer hoppings are regulated by the diameters $d_1$ and $d_2$ of the blue and red tubes, respectively. Other parameters are $a = 52mm$, $d = 20mm$, $h_1 = 17mm$, $d_1 = 8mm$, $d_2 = 5mm$, and $h_2 = 50mm$. (h) Gapless Wilson loop for the valence bands in (c). Inset illustrates the Wilson loop approach.

To observe the bulk-defect correspondence for fragile topology, we design a bilayer phononic crystal according to the tight-binding model in Fig. 1(a) to realize the fragile topological phase. In air-borne phononic crystals, each lattice site can be emulated by a cylindrical acoustic cavity, as shown in the inset of Fig. 1(g). These cylindrical acoustic cavities have the same geometry. The hoppings between the lattice sites can be realized by tubes connecting these acoustic cavities. The radii of these tubes control the intralayer and interlayer couplings. By choosing proper geometry parameters (see the caption of Fig. 1), we achieve the phononic band structure shown in Fig. 1(g). We consider the four lowest phononic bands, which are divided into the valence and conduction bands by a phononic band gap.

*Theoretical analysis of the fragile topology.* – The phononic band gap has fragile topology which can be revealed in a number of ways. First, we calculate the Wilson loop for the two valence bands. We find that the Wilson loop is gapless, which resembles the Wilson loop of quantum spin Hall insulators. However, the system has neither spin-orbit coupling (SOC) nor Kramers degeneracy (as it is a bosonic system). Therefore, it cannot be a 2D topological insulator, but instead a fragile topological insulator protected by the $\pi$-rotation around the $z$ axis and the time-reversal symmetry ($C_2T$) [1, 54]. Tight-binding calculation shows that the system supports gapped edge states and in-gap corner states. In the acoustic realization, due to the chiral symmetry breaking, the corner states disappear. These features confirm the nature of fragile topology [4, 6, 7, 12, 17].

The fragile band topology can be revealed more directly via the band representations based on the theory of topological quantum chemistry (TQC) [2, 5, 14, 16, 43, 44]. The fundamental building blocks in the TQC are the so-called elementary band representations (EBRs). Each EBR corresponds to an atomic limit originating from minimal symmetric Wannier orbitals localized at certain maximal Wyckoff positions, i.e., the high-symmetry points in the unit cell. Remarkably, these atomic limits induce distinct symmetry representations of Bloch eigenstates at the high-symmetry points in the momentum space, indicating a strong connection between the real- and momentum-space descriptions. Most importantly, the attempt to characterize the topological bands in terms of EBRs leads to the classification of trivial atomic insulators, obstructed atomic insulators, stable band topology, and fragile band topology [1, 2]. Trivial and obstructed atomic insulators can always be expressed by the summation of a number of EBRs. In contrast, fragile and stable topological phases cannot be expressed by the summation of a number of EBRs. Fragile topology is often diagnosed by the subtraction of some EBRs from a number of other EBRs, beside the gapless Wilson loop.

The bilayer lattice has the $P_6$ wallpaper symmetry, admitting three types of maximal Wyckoff positions [55], as labeled by $1a$, $2b$, and $3c$ in the left-top panel in Table. 1. The site-symmetry groups at $1a$, $2b$, and $3c$ are isomorphic to the rotation symmetries $C_6$, $C_3$, and $C_2$, respectively.

The corresponding character tables are depicted in Tables. 1(I-III) [55]. The Wannier orbitals localized at Wyckoff positions are subject to the symmetric representations of the corresponding site-symmetry groups, leading to a total of eight EBRs, as summarized in the first two rows in Table. 1(IV). These eight EBRs give rise to distinct symmetry representations $\#_n$ in momentum space, where # denotes the high-symmetry points $\Gamma$, $K$, and $M$, whose symmetry groups are also isomorphic to the rotation symmetries $C_6$, $C_3$, and $C_2$, respectively. The subscript $n$ represents the index of symmetry representations, corresponding to the first column in each character table. Based on the symmetry properties of the bands in Fig. 1(b), we find that the band representation of the bilayer honeycomb lattice with vanishing interlayer hoppings is denoted by $2(A)_{2b} \uparrow G$, as highlighted by the black outline in Table. 1(IV). Once the spiral interlayer hoppings are introduced, a band gap emerges [see Fig. (1c)], leading to the split band representations. Specifically, based on the momentum-space symmetry representations labeled in Fig. 1(c), the split band representations cannot be expressed by the summation of certain EBRs, but have only the correspondence to the combination of both the summation and subtraction, i.e.,

$$(A)_{1a} \uparrow G \oplus (^1E_2\,^2E_2)_{1a} \uparrow G \oplus (A)_{3c} \uparrow G \ominus (^1E\,^2E)_{2b} \uparrow G \tag{1}$$

and

$$(B)_{1a} \uparrow G \oplus (^1E_1\,^2E_1)_{1a} \uparrow G \oplus (B)_{3c} \uparrow G \ominus (^1E\,^2E)_{2b} \uparrow G \tag{2}$$

for the conduction and valence bands, respectively. The inevitable subtraction in band representations explicitly indicates the fragile band topology.

The fragile band topology can be diagnosed by the presence of gapless spectral flows induced by the $C_6$-symmetric screw dislocation, which, essentially, is determined by the nonzero real-space topological invariants (RSTIs). According to Ref. [45], the so-called RSTIs for a $C_6$-symmetric finite lattice with its center being at the Wyckoff position $1a$ is defined by

$$\delta_1 = -m(A) + m(^1E_2\,^2E_2)$$

$$\delta_2 = -m(A) + m(^1E_1\,^2E_1)$$

$$\delta_3 = -m(A) + m(B), \tag{3}$$

where $m$ denotes the multiplicity of the symmetry representations. The RSTIs capture the difference among the number of bulk states with distinct symmetry representations in a finite $C_6$-symmetric system. Following Ref. [45], the RSTIs can be deduced more directly from the symmetry representations in momentum space,

$$\delta_1 = \frac{1}{2}m(\Gamma_2) + m(\Gamma_3\Gamma_5) + m(\Gamma_4\Gamma_6) + m(K_2K_3) - m(M_1) - \frac{1}{2}m(M_2)$$

$$\delta_2 = m(\Gamma_2) + 2m(\Gamma_3\Gamma_5) + m(\Gamma_4\Gamma_6) + m(K_2K_3) - m(M_1) - m(M_2)$$

$$\delta_3 = \frac{3}{2}m(\Gamma_2) + 2m(\Gamma_3\Gamma_5) + m(\Gamma_4\Gamma_6) - m(M_1) - \frac{1}{2}m(M_2). \tag{4}$$

Based on the symmetry representations labeled in Fig. 1(c), we obtain the RSTIs as $\delta_1 = 0$ and $\delta_2 = \delta_3 = -1$. The nonzero RSTIs indicate the presence of three additional states (modulo six) with symmetry representations $A$ and $^1E_2{}^2E_2$ for the conduction bands ($B$ and $^1E_1{}^2E_1$ for the valence bands) as the fundamental feature of filling anomaly. By varying the dislocation-induced artificial gauge flux from 0 to $2\pi$, these additional states with different $C_6$ eigenvalues transform cyclically among one another ($^1E_2 \to A \to {}^2E_2 \to {}^1E_1 \to B \to {}^2E_1 \to {}^1E_2$) and traverse the full band gap to recover the gauge-invariant band structure, manifesting the gapless spectral flows as the signature of fragile topology (see Supplementary Materials for details).

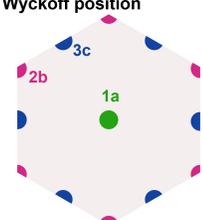

TABLE. 1. Top-left panel: Illustration of the maximal Wyckoff positions $1a$, $2b$, and $3c$ for the $P_6$ wallpaper group. (I)-(III) Character tables of the rotation symmetries $C_2$, $C_3$, and $C_6$. The first and second columns represent the notations of symmetry representations in momentum and real space, respectively. $\epsilon$

($\sigma$) denotes $2\pi/3$ ($2\pi/6$). (IV) The elementary band representations for the $P_6$ wallpaper group in the presence of TRS and in the absence of spin degree of freedom. The first two rows denote the Wannier orbitals with specific symmetry representations localized at certain Wyckoff positions. The other three rows give the corresponding symmetry representations at high-symmetry points $\Gamma$, $K$, and $M$ in momentum space. The outlined rectangle highlights the band representation of the honeycomb lattice.

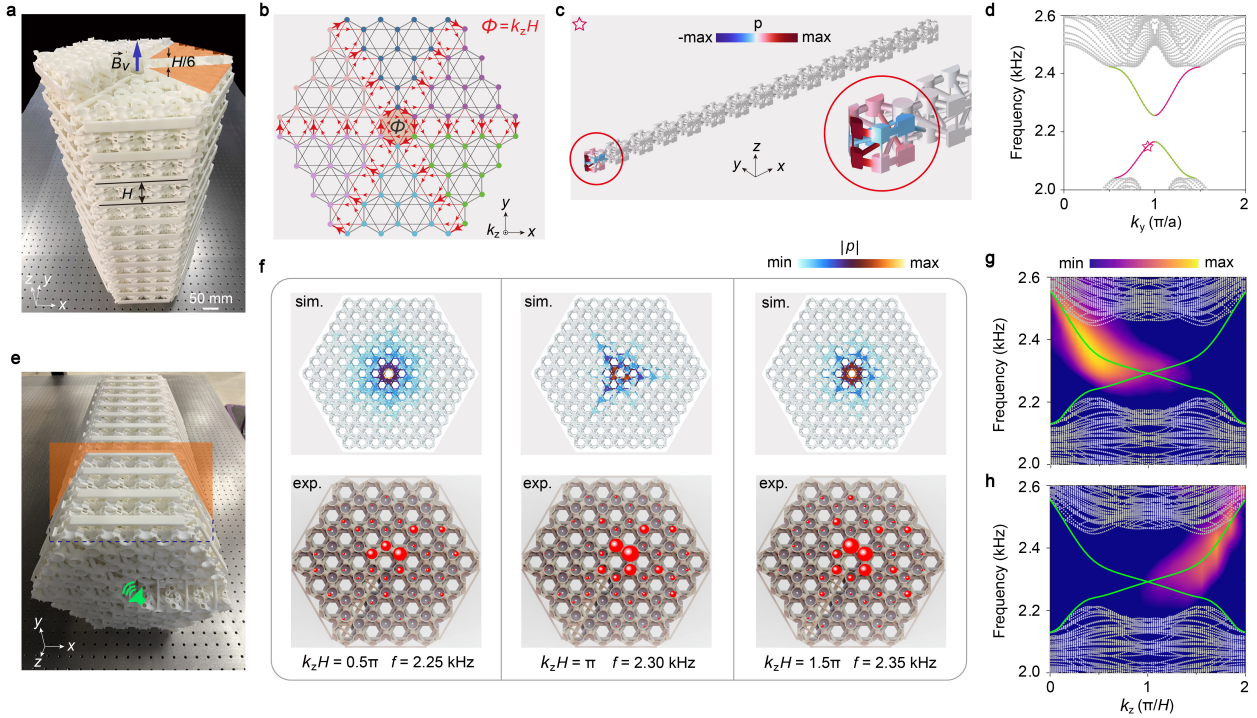

FIG. 2. (a) Photograph of the three-dimensional acoustic crystal sample with a step screw dislocation, which exhibits a height difference of $H/6$ between adjacent sectors. $H = 54mm$ denotes the lattice constant along the $z$ direction. (b) Tight-binding representation of the acoustic crystal in (a) after Fourier transformation along the $z$ direction. The red arrows represent the complex-valued hoppings. Gauge flux insertion at the central plaquette (orange) is indicated. (c) The acoustic pressure field profile of a representative edge state is marked by a pentacle in (d). The inset shows the zoomed-in field pattern. (d) The projected band structure calculated from a ribbon-like supercell with the open (periodic) boundary condition in the $x$ ($y$) direction, which displays the gapped edge states. (e) The location of the loudspeaker for detecting the localized dislocation states, as well as the dispersion. The orange zone denotes the plane where the microphone detects the dislocation states presented in (f). Representative localized dislocation states from both simulation and experiments, showing a good agreement with each other. The radii of red spheres are proportional to the amplitudes of experimentally detected pressure fields. The corresponding Bloch

momenta $k_z$ and frequencies are labeled below each figure. (g) and (h) Experimentally measured spectral flows when the loudspeaker is placed at the center of the top and bottom layers, respectively.

*Experiments.* – We fabricate the phononic crystal with a screw dislocation using 3D printing technology based on photosensitive resins. A photo of the fabricated sample is shown in Fig. 2(a). To make the fabrication easier, we adopt a design of step screw dislocation where the screw is divided into six steps. Each step takes an elevation of *H*/6 along the z direction. Thus, the system has six connected sectors, and the connections between adjacent sectors are tilted. The step screw dislocation induces exactly the same local artificial gauge flux as in a continuous screw dislocation [48, 49]. In fact, from the tight-binding picture and the dimensional reduction, these two screw dislocations can be connected by a gauge transformation [see the effective tight-binding models after dimensional reduction in Fig. 2(b) and see more in Supplemental Materials]. We remark that the gauge phases in the effective tight-binding models after dimensional reduction emerge due to the spiral geometry around the dislocation core (a key feature of the screw dislocation): At a given $k_z$ any translation along the z direction will pick up a phase according to the Bloch theorem. Therefore, in the step screw dislocation hoppings across different sectors pick up a phase $k_z H/6$ along the red arrows in Fig. 2(b). Similarly, in the continuous screw dislocation, nearly all hoppings pick up some phases, according to the Bloch theorem. Nevertheless, in both cases, only the central plaquette has a nonvanishing artificial gauge flux of $\phi = k_z H$. This local artificial gauge flux is the key origin of the bulk-defect correspondence discovered here. Therefore, we expect the same bulk-defect response of the fragile topology emerging as well in the phononic crystal with the step screw dislocation, which is confirmed in both tight-binding calculations and acoustic simulations [see Supplemental Materials].

Before looking at the bulk-defect response, we first check the conventional bulk-edge response in the fragile topological phononic crystal. We calculate the phononic spectrum for a ribbon-like supercell, which is finite in the *x* direction but periodic in the *y* direction [Fig. 2(c)]. We find that the system indeed has gapped edge states [Fig. 2(d)] despite the fact that the Wilson loop of the valence bands is gapless, which is a smoking-gun signature of the fragile topology.

The signature of the bulk-defect correspondence of fragile topology here is the emergence of the 1D gapless bound states at the screw dislocation. In the experiments, we verify the existence of such states via two different probes. First, we measure the wave function of these topological bound states. By placing an acoustic source at one end of the screw dislocation and setting the excitation frequency nearly in the middle of the acoustic band gap, we can probe the acoustic wave pattern of the topological bound states. As the acoustic source excites such topological bound states, they propagate from one end of the screw dislocation to the other end. Using a tiny microphone, we can detect the acoustic wave pattern of these topological bound states at an intersection plane (the detection plane), as illustrated in Fig. 2(e) [see more details in Supplemental Materials]. By measuring the acoustic wave patterns at different detection planes (in fact, we detect the acoustic pressure field across the entire phononic crystal), we obtain the 3D wavefunction of the topological bound states for various excitation frequencies. We further perform a Fourier transformation along the z direction to extract the wavefunction of the topological bound states in the 2D detection plane with varying wavevector $k_z$ and frequency. Several such wavefunctions are presented in Figs. 2(f-h). It is seen that the wavefunctions are indeed localized at the screw dislocation and agree excellently with the simulated wavefunctions at the same wavevector $k_z$ and frequency that are obtained by solving the acoustic eigenstates via finite-element methods.

The above measurements also give the dispersion of the 1D topological bound states at the screw dislocation. As shown in Figs. 2(i-j), the measured dispersion of the 1D topological bound states agrees well with the simulated dispersions. In these measurements, we adopted two pump-probe configurations. In the measurements for Fig. 2(i), the acoustic source is placed at one end of the screw dislocation, as shown in Fig. 2(e), while in the measurements for Fig. 2(j), the acoustic source is placed at the other end of the screw dislocation. The former (latter) configuration excites acoustic waves with positive (negative) group velocity along the z direction. These data agree with the simulated eigenstate spectrum, which clearly shows that the 1D topological bound states are gapless and traverse the entire phononic band gap.

*Conclusion and outlook.* – In this work, we devised a phononic system to uncover an intriguing property of fragile topology: the emergence of 1D gapless topological bound modes at a screw dislocation. With consistent theory, simulations, and experiments, we unveil a novel topological bulk-defect correspondence for fragile topological phases that can be used in the experimental identification of fragile topological materials. As this bulk-defect correspondence is an intrinsic property of fragile topological phases and robust against disorder (when the symmetry underlying the fragile topology is preserved, see Supplemental Material), it provides a powerful experimental tool for the characterization of fragile topological phases and materials. In particular, as they are ubiquitous in many solid-state materials, screw dislocations can be exploited for the experimental study of fragile topological materials (e.g., via scanning tunneling microscope measurements), which are by far still missing and demanded for the test of recent theoretical predictions [38-42, 45].

# References

[1] H. C. Po, H. Watanabe, and A. Vishwanath, Fragile Topology and Wannier Obstructions, *Phys. Rev. Lett.* **121**, 126402 (2018)

[2] B. Bradlyn, L. Elcoro, J. Cano, M. G. Vergoiory, Z. Wang, C. Felser, M. L. Aroyo, and B. A. Bernevig, Topological quantum chemistry, *Nature* **547**, 298-305 (2017).

[3] L. Zou, H. C. Po, A. Vishwanath, and T. Senthil, Band structure of twisted bilayer graphene: Emergent symmetries, commensurate approximants, and Wannier obstructions, *Phys. Rev. B* **98**, 085435 (2018).

[4] B. J. Wieder, and B. A. Bernevig, The axion insulator as a pump of fragile topology, arXiv:1810.02373 (2018).

[5] A. Bouhon, A. M. Black-Schaffer, and R.-J. Slager, Wilson loop approach to fragile topology of split elementary band representations and topological crystalline insulators with time-reversal symmetry, *Phys. Rev. B* **100**, 195135 (2019).

[6] Y. Hwang, J. Ahn, and B.-J. Yang, Fragile topology protected by inversion symmetry: Diagnosis, bulk-boundary correspondence, and Wilson loop, *Phys. Rev. B* **100**, 205126 (2019).

[7] S. Liu, A. Vishwanath, and E. Khalaf, Shift Insulators: Rotation-Protected Two-Dimensional Topological Crystalline Insulators, *Phys. Rev. X* **9**, 031003 (2019).

[8] M. B. de Paz, M. G. Vergniory, D. Bercioux, A. García-Etxarri, and B. Bradlyn, Engineering fragile topology in photonic crystals: Topological quantum chemistry of light, *Phys. Rev. Res*. **1**, 032005(R) (2019).

[9] S. H. Kooi, G. van Miert, and C. Ortix, Classification of crystalline insulators without symmetry indicators: Atomic and fragile topological phases in twofold rotation symmetric systems, *Phys. Rev. B* **100**, 115160 (2019).

[10] Z. Wang, B. J. Wieder, J. Li, B. Yan, and B. A. Bernevig, Higher-Order Topology, Monopole Nodal Lines, and the Origin of Large Fermi Arcs in Transition Metal Dichalcogenides XTe2 (X =Mo,W), *Phys. Rev. Lett* **123**, 186401 (2019).

[11] J. Ahn, and B.-J. Yang, Symmetry representation approach to topological invariants in $C_{2z}T$-symmetric systems, *Phys. Rev. B* **99**, 235125 (2019).

[12] W. A. Benalcazar, T. Li and T. L. Hughes, Quantization of fractional corner charge in $C_n$-symmetric higher-order topological crystalline insulators, *Phys. Rev. B* **99**, 245151 (2019).

[13] H.-X. Wang, G.-Y. Guo, and J.-H. Jiang, Band topology in classical waves: Wilson-loop approach to topological numbers and fragile topology, *New Journal of Physics*, **21**, 093029 (2019).

[14] B. Bradlyn, Z. Wang, J. Cano and B. A. Bernevig, Disconnected elementary band representations, fragile topology, and Wilson loops as topological indices: An example on the triangular lattice, *Phys. Rev. B* **99**, 045140 (2019).

[15] Z.-D. Song, L. Elcoro, Y.-F. Xu, N. Regnault, and B. A. Bernevig, Fragile Phases as Affine Monoids: Classification and Material Examples, *Phys. Rev. X* **10**, 031001 (2020).

[16] A. Alexandradinata, J. Höller, C. Wang, H. Cheng, and L. Lu, Crystallographic splitting theorem for band representations and fragile topological photonic crystals, *Phys. Rev. B* **102**, 115117 (2020).

[17] C. Shang, X. Zang, W. Gao, U. Schwingenschlögl, and A. Manchon, Second-order topological insulator and fragile topology in topological circuitry simulation, arXiv:2009.09167 (2020).

[18] J. L. Mañes, Fragile phonon topology on the honeycomb lattice with time-reversal symmetry, *Phys. Rev. B* **102**, 024307 (2020).

[19] S. Kobayashi, and A. Furusaki, Fragile topological insulators protected by rotation symmetry without spin-orbit coupling, *Phys. Rev. B* **104**, 195114 (2021).

[20] R.-X. Zhang, and Z.-C. Yang, Tunable fragile topology in Floquet systems, *Phys. Rev. B* **103**, L121115 (2021).

[21] Y.-F. Chen, and D.-X. Yao, Fragile topological phase on the triangular kagome lattice and its bulk-boundary correspondence, *Phys. Rev. B* **107**, 155129 (2023).

[22] G. F. Lange, A. Boulun, and R.-J. Slager, Spin texture as a bulk indicator of fragile topology, *Phys. Rev. Res.* **5**, 033013 (2023).

[23] H. C. Po, L. Zou, T. Senthil, and A. Vishwanath, Faithful tight-binding models and fragile topology of magic-angle bilayer graphene, *Phys. Rev. B* **99**, 195455 (2019).

[24] J. Ahn, S. Park, and B.-J. Yang, Failure of Nielsen-Ninomiya theorem and fragile topology in two dimensional systems with space-time inversion symmetry: Application to twisted bilayer graphene at magic angle, *Phys. Rev. X* **9**, 021013 (2019).

[25] Z.-D. Song, Z. Wang, W. Shi, G. Li, C. Fang, and B. A. Bernevig, All magic angles in twisted bilayer graphene are topological, *Phys. Rev. Lett.* **123**, 036401 (2019).

[26] B. Lian, F. Xie, and B. A. Bernevig, Landau level of fragile topology, *Phys. Rev. B* **102**, 041402(R) (2020).

[27] Z.-D. Song, B. Lian, N. Regnault, and B. A. Bernevig, Twisted bilayer graphene. II. Stable symmetry anomaly, *Phys. Rev. B* **103**, 205412 (2021).

[28] J. Ahn, D. Kim, Y. Kim, and B.-J. Yang, Band Topology and Linking Structure of Nodal Line Semimetals with $Z_2$ Monopole Charges, *Phys. Rev. Lett*. **121**, 106403 (2018).

[29] F. N. Ünal, A. Bouhon, and R. J. Slager, Topological Euler class as a dynamical observable in optical lattices, *Phys. Rev. Lett*. **125**, 053601 (2020).

[30] A. Tiwari, and T. Bzdušek, Non-Abelian topology of nodal-line rings in *PT*-symmetric systems, *Phys. Rev. B* **101**, 195130 (2020).

[31] A. Bouhon, Q. Wu, R.-J. Slager, H. Weng, O. V. Yazyev, and T. Bzdušek, Non-Abelian reciprocal braiding of Weyl points and its manifestation in ZrTe, *Nature Physics* **16**, 1137–1143 (2020).

[32] A. Bouhon, T. Bzdušek, and R.-J. Slager, Geometric approach to fragile topology beyond symmetry indicators, *Phys. Rev. B* **102**, 115135 (2020).

[33] Y. Guan, A. Bouhon, and O. V. Yazyev, Landau levels of the Euler class topology, *Phys. Rev. Res.* **4**, 023188 (2022).

[34] B. Jiang, A. Bouhon, S.-Q. Wu, Z.-L. Kong, Z.-K. Lin, R.-J. Slager, and J.-H. Jiang, Experimental observation of meronic topological acoustic Euler insulators, arXiv:2205.03429v1 (2022).

[35] C. S. Chiu, D.-S. Ma, Z.-D. Song, B. A. Bernevig, and A. A. Houck, Fragile topology in line-graph lattices with two, three, or four gapped flat bands, *Phys. Rev. Res.* **2**, 043414 (2020).

[36] A. Skurativska, S. S. Tsirkin, F. D. Natterer, T. Neupert, and M. H. Fischer, Flat bands with fragile topology through superlattice engineering on single-layer graphene, *Phys. Rev. Res.* **3**, L032003 (2021).

[37] D. Călugăru, A. Chew, L. Elcoro, Y. Xu, N. Regnault, Z.-D. Song, and B. A. Bernevig, General construction and topological classification of crystalline flat bands, *Nature Physics* **18**, 185-189 (2022).

[38] S. Peotta, and P. Törmä, Superfluidity in Topologically Nontrivial Flat Bands, *Nat. Commun.* **6**, 8944 (2015).

[39] F. Xie, Z. Song, B. Lian, and B. A. Bernevig, Topology-Bounded Superfluid Weight in Twisted Bilayer Graphene, *Phys. Rev. Lett.* **124**, 167002 (2020).

[40] X. Wang, and T. Zhou, Fragile topology in nodal-line semimetal superconductors, *New Journal of Physics* **24**, 083013 (2022).

[41] V. Peri, Z.-D. Song, B. A. Bernevig, and S. D. Huber, Fragile topology and flat-band superconductivity in the strong-coupling regime, *Phys. Rev. Lett.* **126**, 027002 (2021).

[42] P. Törmä, S. Peotta, and B. A. Bernevig, Superconductivity, superfluidity and quantum geometry in twisted multilayer systems, *Nature Reviews Physics* **4**, 528-542 (2022).

[43] J. Cano, B. Bradlyn, Z. Wang, L. Elcoro, M. G. Vergniory, C. Felser, M. I. Aroyo and B. A. Bernevig, Building blocks of topological quantum chemistry: Elementary band representations, *Phys. Rev. B* **97**, 035139 (2018).

[44] J. Cano, B. Bradlyn, Z. Wang, L. Elcoro, M. G. Vergniory, C. Felser, M. I. Aroyo and B. A. Bernevig, Topology of Disconnected Elementary Band Representations, *Phys. Rev. Lett.* **120**, 266401 (2018).

[45] Z.-D. Song, L. Elcoro, and B. A. Bernevig, Twisted bulk-boundary correspondence of fragile topology, *Science* **367**, 794-797 (2020).

[46] V. Peri, Z. Song, M. Serra-Garcia, P. Engeler, R. Queiroz, X. Huang, W. Deng, Z. Liu, B. A. Bernevig, and S. D. Huber, Experimental characterization of fragile topology in an acoustic metamaterial, *Science* **367**, 797-800 (2020).

[47] M. Miniaci, F. Allein, and R. K. Pal, Spectral flow of a localized mode in elastic media, arXiv:2111.09021v1 (2021).

[48] Z.-K. Lin, Y. Wu, B. Jiang, Y. Liu, S.-Q. Wu, F. Li, and J.-H. Jiang, Topological Wannier cycles induced by sub-unit-cell artificial gauge flux in a sonic crystal, *Nature Materials* **21**, 430-437 (2022).

[49] Z.-L. Kong, Z.-K. Lin, and J.-H. Jiang, Topological Wannier Cycles for the Bulk and Edges, *Chinese Physics Letters* **39**, 084301 (2022).

[50] B. Xie, W. Deng, J. Lu, H. Liu, P. Lai, H. Cheng, Z. Liu, and S. Chen, Correspondence between real-space topology and spectral flows at disclinations, *Phys. Rev. B* **108**, 134118 (2023).

[51] G. Palumbo, Non-Abelian Tensor Berry Connections in Multiband Topological Systems, *Phys. Rev. Lett.* **126**, 246801 (2021).

[52] A. Bouhon and R.-J. Slager, Multi-gap topological conversion of Euler class via band-node braiding: minimal models, PT-linked nodal rings, and chiral heirs, arXiv:2203.16741 (2022).

[53] R. Takahashi and T. Ozawa, Bulk-edge correspondence of Stiefel-Whitney and Euler insulators through the entanglement spectrum and cutting procedure, *Phys. Rev. B* **108**, 075129 (2023).

[54] C. Brouder, G. Panati, M. Calandra, C. Mourougane, and N. Marzari, Exponential localization of Wannier functions in insulators, *Phys. Rev. Lett.* **98**, 046402 (2007).

[55] L. Elcoro, B. Bradlyn, Z. Wang, M. G. Vergniory, J. Cano, C. Felser, B. A. Bernevig, D. Orobengoa, G. de la Flor, and M. I. Aroyo, Double crystallographic groups and their representations on the Bilbao Crystallographic Server, *J. Appl. Cryst.* **50**, 1457–1477 (2017).

# Acknowledgements

J.-H. J. thanks the supports from the National Key R&D Program of China (2022YFA1404400), the National Natural Science Foundation of China (Grant Nos. 12125504 and 12074281), the "Hundred Talents Program" of the Chinese Academy of Sciences, and the Priority Academic Program Development (PAPD) of Jiangsu Higher Education Institutions. Z.-D. S. was supported by the National Natural Science Foundation of China (General Program No. 12274005), National Key R&D Program of China (No. 2021YFA1401900). Y. W. was supported by the Fundamental Research Funds for the Central Universities (No. 30923010207) and the National Natural Science Foundation of China (Grant No. 12302112).